\documentclass[10pt,journal]{IEEEtran}

\usepackage{amsmath,amssymb,epsfig,nicefrac,color}
\usepackage{amsfonts,graphics,subfigure,epsfig,graphics,cleveref,url}

\usepackage{mathrsfs}
\usepackage{cite,calc}

\newcommand\nc\newcommand
\nc\bfa{{\boldsymbol a}}\nc\bfA{{\boldsymbol A}}\nc\cA{{\mathcal A}}
\nc\bfb{{\boldsymbol b}}\nc\bfB{{\boldsymbol B}}\nc\cB{{\mathcal B}}
\nc\bfc{{\boldsymbol c}}\nc\bfC{{\boldsymbol C}}\nc\cC{{\mathcal C}}
\nc\sC{{\mathscr C}}
\nc\bfd{{\boldsymbol d}}\nc\bfD{{\boldsymbol D}}\nc\cD{{\mathcal D}}
\nc\bfe{{\boldsymbol e}}\nc\bfE{{\boldsymbol E}}\nc\cE{{\mathcal E}}
\nc\bff{{\boldsymbol f}}\nc\bfF{{\boldsymbol F}}\nc\cF{{\mathcal F}}
\nc\bfg{{\boldsymbol g}}\nc\bfG{{\boldsymbol G}}\nc\cG{{\mathcal G}}
\nc\bfh{{\boldsymbol h}}\nc\bfH{{\boldsymbol H}}\nc\cH{{\mathcal H}}
\nc\bfi{{\boldsymbol i}}\nc\bfI{{\boldsymbol I}}\nc\cI{{\mathcal I}}
\nc\bfj{{\boldsymbol j}}\nc\bfJ{{\boldsymbol J}}\nc\cJ{{\mathcal J}}
\nc\bfk{{\boldsymbol k}}\nc\bfK{{\boldsymbol K}}\nc\cK{{\mathcal K}}
\nc\bfl{{\boldsymbol l}}\nc\bfL{{\boldsymbol L}}\nc\cL{{\mathcal L}}
\nc\bfm{{\boldsymbol m}}\nc\bfM{{\boldsymbol M}}\nc\sM{{\mathscr M}}
\nc\bfn{{\boldsymbol n}}\nc\bfN{{\boldsymbol N}}\nc\cN{{\mathcal N}}
\nc\bfo{{\boldsymbol o}}\nc\bfO{{\boldsymbol O}}\nc\cO{{\mathcal O}}
\nc\bfp{{\boldsymbol p}}\nc\bfP{{\boldsymbol P}}\nc\cP{{\mathcal P}}
\nc\bfq{{\boldsymbol q}}\nc\bfQ{{\boldsymbol Q}}\nc\cQ{{\mathcal Q}}
\nc\bfr{{\boldsymbol r}}\nc\bfR{{\boldsymbol R}}\nc\cR{{\mathcal R}}
\nc\bfs{{\boldsymbol s}}\nc\bfS{{\boldsymbol S}}\nc\cS{{\mathcal S}}
\nc\bft{{\boldsymbol t}}\nc\bfT{{\boldsymbol T}}\nc\cT{{\mathcal T}}
\nc\bfu{{\boldsymbol u}}\nc\bfU{{\boldsymbol U}}\nc\cU{{\mathcal U}}
\nc\bfv{{\boldsymbol v}}\nc\bfV{{\boldsymbol V}}\nc\cV{{\mathcal V}}
\nc\bfw{{\boldsymbol w}}\nc\bfW{{\boldsymbol W}}\nc\cW{{\mathcal W}}
\nc\bfx{{\boldsymbol x}}\nc\bfX{{\boldsymbol X}}\nc\cX{{\mathcal X}}
\nc\bfy{{\boldsymbol y}}\nc\bfY{{\boldsymbol Y}}\nc\cY{{\mathcal Y}}
\nc\bfz{{\boldsymbol z}}\nc\bfZ{{\boldsymbol Z}}\nc\cZ{{\mathcal Z}}

\nc{\remove}[1]{}

\DeclareMathOperator{\supp}{supp}

\def\h_q{\qopname\relax{no}{h_q}}
\def\dist{d_H}
\def\avg{{\mathbb E}}
\newtheorem{theorem}{Theorem}
\newtheorem{definition}{Definition}
\newtheorem{lemma}{Lemma}
\newtheorem{proposition}{Proposition}
\newtheorem{corollary}{Corollary}

\newtheorem{remark}{\indent Remark}

\newcommand\ff{{\mathbb F}}

\allowdisplaybreaks

\begin{document}
\sloppy

\title{Nonadaptive Group Testing with Random Set of Defectives}
\author{Arya Mazumdar 
\thanks{The author is with the College of Information and Computer Sciences, University of Massachusetts Amherst, MA 01002, email: \texttt{arya@cs.umass.edu}. 
  A part of this work was presented in the International Symposium on Algorithms and Computation, 2012 \cite{mazumdar2012almost} and took place while the author was in Massachusetts Institute of Technology.
This  research is  supported in part by NSF CCF-1318093, NSF CCF-1642658, and NSF CCF-1642550. 
}}

\allowdisplaybreaks
\maketitle

\begin{abstract}
 In a \emph{group testing} scheme, a series of {\em tests} are designed
to identify a small number $t$ of defective items that are present among a large number $N$ of items.
Each test takes as input a  group of items and produces a binary output indicating whether any defective item is present in the group.
In a non-adaptive scheme  the  tests have to be designed
in one-shot. In this setting, designing a testing scheme is equivalent
to the construction of  a \emph{disjunct matrix}, an $M \times N$ binary
matrix where the union of supports of any $t$ columns does not contain the support of any other column.
In principle, one wants to have such a matrix with minimum possible number $M$ of rows.

In this paper we consider the scenario where defective items are random and follow simple probability distributions.
In particular we consider the cases where 1) each item can be defective independently with probability $\frac{t}{N}$ and
2) each $t$-set of items can be defective with uniform probability. In both cases our aim is to design a testing matrix that
successfully identifies the set of defectives with high probability. Both of these models have been studied in the literature
before and it is known that $\Theta(t\log N)$ tests are necessary as well as sufficient (via random coding) in both cases.

Our main focus is explicit deterministic construction of the test matrices amenable to above scenarios. 
One of the most popular ways of constructing test matrices relies on \emph{constant-weight
error-correcting codes} and their \emph{minimum distance}. 
In particular, it is known that codes result in test matrices with $O(t^2 \log N)$ rows that identify any $t$ defectives. 
We go beyond the minimum distance analysis and 
connect the \emph{average distance} of a constant
weight code to the parameters of the resulting test matrix. 
Indeed, we show how distance, a pairwise property of the columns of the matrix,
translates to a $(t+1)$-wise property of the columns.
With our relaxed requirements, we show that using  explicit constant-weight codes (e.g., based on algebraic geometry codes)
we may achieve a number of tests equal to $O(t \frac{\log^2 N}{ \log t})$ for both  the first and the second cases. While only away by a factor of $\frac{\log N}{\log t}$  from the optimal number of tests, this is the best set of parameters one can obtain from a deterministic construction and
 our main contribution lies in relating the group testing properties to
average and minimum distances  of constant-weight codes.  
  \end{abstract}
  
  \begin{IEEEkeywords}
 Group testing, Disjunct matrices, Constant-weight codes, Deterministic construction 
  \end{IEEEkeywords}

\section{Introduction}
Combinatorial  search is an old and well-studied problem.
In the most general form it is assumed that there is a set of $N$ elements
among which at most $t$ are {\em defective}.
This set of defective items is called the {\em defective set} or {\em configuration.}
To find the defective set, one might
test all the elements individually  for defects, requiring $N$ tests. Intuitively, that would be a waste of resource if
$t \ll N$. On the other hand, to identify the defective configuration it is required to ask at least
$\log \sum_{i=0}^t \binom{N}{i} \approx t \log \frac{N}{t}$  yes-no questions. The main objective is to identify the defective
configuration with a number of tests that is as close to this minimum as possible.

In the {\em group testing} problem, a {\em group} of elements are tested together  and
if this particular group contains any defective
element the test result is positive. Based on the test results of this kind one {\em identifies} (with an efficient
algorithm) the  defective set with minimum possible number of tests.
The schemes (grouping of elements) can be adaptive, where
the design of one test may depend on the results of preceding tests. For a comprehensive survey
of adaptive group testing schemes we refer the reader to \cite{du1993combinatorial}.

In this paper we are interested in non-adaptive group testing schemes: here
all the tests are designed together. If the number of designed tests is $M$, then a
non-adaptive group testing scheme is equivalent to the design of a  binary {\em test matrix}
of size $M \times N$ where the $(i,j)$th entry is $1$ if the $i$th test includes the $j$th element;
it is $0$ otherwise. As the test results,  we see  the Boolean OR of the
columns corresponding to the defective entries.

Extensive research has been performed to find out the minimum
number of required tests $M$ in terms of the number of elements $N$ and the maximum number of
defective elements $t$. The best known lower bound says that it is necessary
to have $M = \Omega(\frac{t^2}{\log{t}} \log N)$ tests \cite{d1982bounds, dyachkov1989superimposed}.
The existence of non-adaptive group testing
 schemes with $M = O(t^2 \log N)$ is also known for quite some time \cite{du1993combinatorial, hwang1987non}.
 On the other hand, for the adaptive
setting, schemes have been constructed  with as small as $O(t\log N)$ tests,  optimal
up to a constant factor \cite{du1993combinatorial, hwang1972method}.

In the literature, many relaxed versions of the group testing problem have been studied as well.
For example, in \cite{dyach1983survey,vilenkin1998constructions} recovery of a {\em list} of items containing 
 the true defectives is suggested (list-decoding superimposed codes). This notion was revisited in  \cite{cheraghchi2009noise, indyk2010efficiently}
 as list-disjunct matrix and in
\cite{gilbert2008group}, where it was assumed that recovering a large fraction of
defective elements is sufficient. There are also information-theoretic
models for the group testing problem where the test results can be noisy \cite{atia2012boolean} (also see \cite{chan2012non,cai2013grotesque}). 
In other versions of the group testing problem, a test may carry more than
one bit of information \cite{hwang1984three, blumenthal1971symmetric}, or the test results are threshold-based (see \cite{cheraghchi2013improved}
and references therein). Algorithmic aspects of the recovery schemes have
been studied in several papers. For example, papers \cite{indyk2010efficiently} and \cite{ngo2011efficiently} provide
 efficient recovery algorithms for non-adaptive group testing. 

Here as well, we consider two relaxed versions of the group testing problem --  we want recovery to be successful
with high probability assuming uniform distributions of the defective items. In the first scenario,
each of the $N$ items can be defective with probability $\frac{t}{N}$. This model of defectives, called {\bf Model 1} throughout the rest of this paper, is as old as
the group testing problem \cite{dorfman1943detection} and was rigorously defined in \cite{sobel1966bin}. It is also the subject of very recent works such as
\cite{scarlett2016phase}.
We provide explicit construction of test matrices with $O(t\log^2N/\log t)$ tests for this situation.
%
In the second scenario, we want the recovery to be successful for 
a very large fraction of all possible $t$-sets as defective configurations.
This scenario, called {\bf Model 2} throughout this paper, was
 considered under the name of {\em weakly separated design} in
\cite{malyutov1978separating}, \cite{zhigljavsky2003probabilistic} and \cite{macula2004trivial}.
It is known (see, \cite{zhigljavsky2003probabilistic}) that, with this relaxation
it might be possible to reduce the number of tests to be proportional to $t\log N$. However
this result is not constructive.
Here also we provide explicit construction of test matrices with $O(t\log^2N/\log t)$ tests. Note that,
this result is order-optimal when $t$ is proportional to $N^\delta$ for $0 < \delta \le 1$.

In particular, our result leads to improvement over the  construction of weakly-separated design
from  \cite{gilbert2012recovering}, whenever $\log N \le (\log t)^2$. In \cite{gilbert2012recovering}, the total $N$ items are partitioned and then 
a nonadaptive scheme for a smaller set of elements is
 repeated on each of the parts. It follows from a simple union bound that one would need $O(t \log t \log N)$ tests
 for both the above random models to have high probability identification.

The repeated-block construction of \cite{gilbert2012recovering} is analogous to repeating a good error-correcting code of small
length to construct a long error-correcting code. Indeed, one can find the best linear error-correcting code of length $\log n$ and then repeat that
$n/\log n$ times to construct a capacity-achieving code of length $n$. While this can be a  first construction, it does not give any insight 
regarding the properties that are important for the problem. In an earlier conference version \cite{mazumdar2012almost} of this paper, we showed that the properties of the distribution of
Hamming distances of the columns of testing matrix can play a role in identification. While the result of \cite{mazumdar2012almost} leads to suboptimal number of tests,
we can use better concentration inequalities to arrive at improvements over it (see, Theorem~\ref{thm:main2}). Our construction also turns out to give better parameters than the repetition scheme of \cite{gilbert2012recovering}, whenever $\log N \le (\log t)^2$. Note that, this in particular include the regime where $t$ varies as $N^\delta$ for $0<\delta <1$,  which is the premise of very recent works such as Scarlett and Cevher \cite{scarlett2016phase}. 
There is no apparent relation to the work of \cite{gilbert2012recovering} with our techniques. 
In particular, our ideas cannot be viewed as an extension of repeated block construction.

We believe that our main contribution  lies in 1) relating the group testing properties to the average Hamming distance between the columns of testing matrix and 2) using proper classes of explicit codes (such as Algebraic-Geometric codes) that satisfy the required properties of average and minimum distances.

Non-adaptive group testing has found applications
in multiple different areas, such as, multi-user communication \cite{berger1984random, wolf1985born}, DNA screening \cite{ngo2000survey},
pattern finding \cite{macula2004group} etc.
It can be observed that in many of these applications it would have been still useful
to have a scheme that identifies almost all different defective configurations if not
all possible defective configurations. 
The above relaxations  form a parallel of similar works in 
 compressive sensing (see, \cite{calderbank2010construction,mazumdar2011sparse})
where recovery of almost all sparse signals from a
generic random model is considered.

A construction of group testing schemes from error-correcting code matrices and using code concatenation
appeared in the seminal paper by Kautz and Singleton \cite{kautz1964nonrandom}.
Code concatenation is a way to construct binary codes from codes over a larger alphabet \cite{MS1977}.
In \cite{kautz1964nonrandom}, the authors
concatenate a $q$-ary ($q>2$) Reed-Solomon code with a unit-weight code to use the resulting codewords as the columns of the testing matrix.
Recently in \cite{porat2008explicit}, an explicit construction of a scheme with $M = O(t^2 \log N)$ tests
is provided. The construction of \cite{porat2008explicit} is based on the idea of \cite{kautz1964nonrandom}: instead of
the Reed-Solomon code, they take a low-rate code that achieves the Gilbert-Varshamov bound
of coding theory \cite{MS1977, roth2006introduction}. Papers, such as \cite{d2000new,yekhanin1998some}, also consider construction
of non-adaptive group testing schemes.

In this paper we show that the explicit construction of \cite{porat2008explicit} based on error-correcting codes
works for both Model 1 and Model 2 and results in numbers of tests claimed above.
Not only that, using explicit families of Algebraic-Geometric codes in conjunction with Kautz and Singleton construction 
we  obtain test-matrices with the same performance guarantee.

\subsection{Results and organization}
The constructions of \cite{kautz1964nonrandom, porat2008explicit} and many others are based on 
{\em constant-weight error-correcting codes}, a set of binary vectors of same Hamming weight (number of ones).
The group-testing recovery property relies on the pairwise {\em minimum distance}
between the vectors of the code \cite{kautz1964nonrandom}. In this work,  we go beyond this
minimum distance analysis and relate the group-testing parameters to the
{\em average distance} of the constant-weight code. This allows us to connect
the group testing matrices designed for random models of defectives to 
error-correcting codes in a general way (see, Thm.~\ref{thm:indep} and Thm.~\ref{thm:main1}). Previously the connection 
between distances of the code and weakly separated designs
 was only known for the very specific family of {\em maximum distance
separable} codes \cite{macula2004trivial}, where much more information than the
average distance is evident.

Based on the newfound connection, for both Model 1 and Model 2, we construct  explicit (constructible deterministically in polynomial time)
  families  of non-adaptive group testing schemes. This result can be summarized in the following informal theorem.
 \begin{theorem}[Informal] 
  For both Models 1 and 2, our deterministic nonadaptive scheme can identify 
  the set of defectives exactly with probability $1-\epsilon$. The sufficient number of  tests required for this is $O(t \frac{\ln N}{\ln t} \ln \frac{N}{\epsilon})$.
  \end{theorem}

One of  our construction technique
is same as the scheme of \cite{kautz1964nonrandom,porat2008explicit}, however with a finer analysis
relying on the distance properties of a linear code
we are able to achieve more. We also use explicit families of Algebraic-Geometric codes to obtain the same set of parameters.
One of the main contribution is to show a general way to establish  a property
 for almost all $t$-tuples of elements from a set based on the mean pairwise statistics
of the set.

In Section \ref{sec:disjunct}, we provide the necessary definitions and preliminaries. 
The relation of group testing parameters of Model 1 with constant-weight codes is provided in 
Section \ref{sec:model1}.
In 
Section  \ref{sec:proof1}
we establish the connection between the parameters of a weakly separated design and the average distance of a 
constant-weight code.
In Section \ref{sec:construction}
we discuss our  construction schemes (including one that relies on Algebraic-Geometric codes) that work for both of Models 1 and 2.

\section{Basic definitions and properties}\label{sec:disjunct}

A vector is denoted by bold lowercase letters, such as $\bfx$, and the $i$th entry of the vector $\bfx$ is denoted by
$x_i$. The Hamming distance between two vectors is denoted by $\dist(\cdot,\cdot)$.
The {\em support} of a vector $\bfx$ is the set of coordinates  where the vector has nonzero
entries. It is denoted by $\supp(\bfx)$. We use the usual set terminology, where a set $A$ contains $B$ if $B\subseteq A.$
Also, below $[n]$ denotes $\{1,2,\dots, n\}$. 

First of all, we define the following two models for {\em defectives}.
\begin{definition}[Random models of defectives]
In the random defective {\em Model 1}, among a set of $N$ elements, each element is independently defective with
probability $\frac{t}{N}.$ In the random defective {\em Model 2},  each subset of cardinality $t$ of a set of $N$ elements has equal probability ${\binom{N}{t}}^{-1}$ of 
being the defective set.
\end{definition}

\subsection{Disjunct matrices}

The following definition of disjunct matrices  is standard and can be found in  \cite[Ch.~7]{du1993combinatorial}.
\begin{definition}
An $M \times N$ binary matrix $A$ is called $t$-disjunct if the support of any column
is not contained in
the union of the supports of any other  $t$ columns.
\end{definition}
It is not very difficult to see that a $t$-disjunct matrix gives a group testing scheme
that identifies any defective set up to size $t$. On the other hand
any group testing scheme
that identifies any defective set up to size $t$ must be a $(t-1)$-disjunct
matrix.
The definition of disjunct matrix can be restated as follows: a matrix is $t$-disjunct if
any  $t+1$ columns  indexed by $i_1,\dots,i_{t+1}$ of the
matrix form a sub matrix which must have a
row that has exactly one $1$ in the  $i_j$th position and zeros in the
other positions, for $j =1, \dots, t+1.$

To a great advantage,  disjunct matrices allow for a simple identification
algorithm that runs in time $O(Nt),$ as we see below.

\subsection{Disjunct decoding}
Given the test results $\bfy \in \{0,1\}^M$, we use the following recovery algorithm to find the defectives. Suppose, $A$ is the 
test matrix and $\alpha^{(j)} \in \{0,1\}^N, j =1, \dots, M$ denotes the $j$th row of $A$. The recovery algorithm simply 
outputs 
$$
[N]\setminus \cup_{j:y_j =0}\supp(\alpha^{(j)} )
$$
as the set of defectives  \cite[Ch.~7]{du1993combinatorial}.

Note that, irrespective of the testing matrix, this algorithm will always output a set that contains all the defective elements. Moreover,
if the testing matrix is disjunct, then the output is exactly equal to the set of defectives.  We have the following simple proposition.

\begin{proposition}\label{prop:disjunct}
Suppose, the set of defectives is $S \subseteq [N]$. Let $\bfa^{(k)}$ denote the $k$th column of the 
test matrix $A$. Then the disjunct decoding algorithm recovers the defectives exactly if
$\cup_{j \in S}\supp(\bfa^{(j)})$ does not contain the support of $\bfa^{(i)}$ for all $i \in [N]\setminus S$.
\end{proposition}

\subsection{Almost disjunct matrices}
Below we define  a {\em relaxed} form of disjunct matrices. This definition appeared very closely in \cite{malyutov1978separating,zhigljavsky2003probabilistic} 
and  exactly in
\cite{macula2004trivial}.
\begin{definition}
For any $\epsilon >0$, an $M \times N$ matrix $A$ is called $(t,\epsilon)$-disjunct if  the set of  $t$-tuple of
columns (of size $\binom{N}t$) has a subset $\cB$ of size at least $(1-\epsilon)\binom{N}t$ with the following property:
for all $J \in \cB$, $\cup_{\kappa\in J}\supp(\kappa)$ does not contain support of any  column $\nu \notin J.$
\end{definition}
In other words, the union of supports of a
randomly and uniformly chosen set of  $t$ columns from a  $(t,\epsilon)$-disjunct matrix does not contain the support of any
other column with probability at least $1-\epsilon$. It is clear that for $\epsilon =0$, the $(t,\epsilon)$-disjunct matrices
are same as $t$-disjunct matrices.

It is easy to see the following fact.
\begin{proposition}[Model 2]\label{prop:scheme}
A  $(t,\epsilon)$-disjunct matrix gives a group testing scheme that can identify
all but at most a fraction $\epsilon >0$ of all possible defective configurations of size  $t$.
\end{proposition}

\subsection{Constant-weight codes}
A binary $(M,N,d)$ code $\cC$ is a set of size $N$ consisting of $\{0,1\}$-vectors of length $M$.
 Here $d$ is the largest integer such that
any two vectors (codewords) of $\cC$ are at least
Hamming distance $d$ apart. $d$ is called the {\em minimum distance} (or {\em distance})
of $\cC.$ If all the codewords of $\cC$ have Hamming weight $w$,
then it is called a constant-weight code. In that case we write $\cC$ is an
$(M,N,d,w)$-constant-weight binary code.

Constant-weight codes can give constructions of group testing schemes.
One just arranges the codewords as the columns of the test matrix. Kautz and Singleton
proved the following in \cite{kautz1964nonrandom}.
\begin{proposition}\label{prop:disj}
An $(M,N,d,w)$-constant-weight binary code provides
a $t$-disjunct matrix where,
$
t = \Big\lfloor\frac{w-1}{w-d/2}\Big\rfloor.
$
\end{proposition}
\begin{IEEEproof}
The intersection of supports of any two columns has size at most $w -d/2$. Hence if
$w> t(w-d/2)$, support of any column will not be contained in the union of supports of any $t$ other columns.
\end{IEEEproof}
Extensions of  Prop.~\ref{prop:disj} are our main results. 
To do that we
need to define the {\em average distance} $D$ of a code $\cC$:
$$
D(\cC) =  \frac{1}{|\cC|} \min_{\bfx \in \cC}\sum_{\bfy \in \cC} \dist(\bfx,\bfy).
$$
Here $\dist(\bfx,\bfy)$ denotes the Hamming distance between $\bfx$ and $\bfy$.
Also define the second-moment of distance distribution:
$$
D_2(\cC) =  \frac{1}{|\cC|^2} \sum_{\bfy,\bfx \in \cC} \dist(\bfx,\bfy)^2.
$$

\section{Model 1: independent defectives - test matrices from constant-weight codes} \label{sec:model1}
In this section, we consider the independent failure model (Model 1) and show how the minimum and average distances of a constant-weight binary code contribute to a nonadaptive group testing scheme.
Recall, in this model we assume that among $N$ items, each is defective with a probability $\frac{t}{N}$. 
The main result of this section is the following theorem.
\begin{theorem}[Model 1]\label{thm:indep}
Suppose, we have a constant-weight binary code $\cC$ of size $N$, minimum distance $d$ and average distance $D$ such that every codeword has length $M$ and weight $w$. The test matrix obtained from the
code exactly identifies all the defective items (chosen according to Model 1)  with probability at least $1-\epsilon$ (over the probability space of Model 1) if 
\begin{equation}
w-\frac{d}{2} \le \frac{3\Big(w - t(w-D/2)\Big)^2}{2 \Big(2t(w-D/2) +w  \Big)\ln \frac{N}{\epsilon}}.
\end{equation}
\end{theorem}

We will need the help of the following lemma to prove the theorem.
Note, from Prop.~\ref{prop:disjunct}, the disjunct-recovery algorithm will be
successful if the union of supports of the columns corresponding to the defectives does not contain the support of any other columns.
Suppose the testing matrix is constructed from an $(M, N, d, w)$-constant-weight code $\cC$ (each column is a codeword). Let
$$
\cC = \{\bfc_1,\bfc_2, \dots, \bfc_N\}.
$$
Moreover, assume $X_j \in \{0,1\}$ is the indicator Bernoulli($t/N$) random variable that denotes whether the $j$th element is defective or not.

\begin{lemma}
Suppose, for all $i \in [N]$, we have 
$$
\sum_{j =1, j \ne i}^N X_j\Big(w - \frac{\dist(\bfc_i,\bfc_j)}2\Big) < w.
$$
Then the disjunct-recovery algorithm will exactly identify the defective elements.
\end{lemma}
\begin{IEEEproof}
The lemma directly follows from Prop.~\ref{prop:disjunct} and the fact that for any $i,j$, $w-\frac{\dist(\bfc_i,\bfc_j)}2$ is nonnegative. Suppose $S\subseteq [N]$ be the random set of defectives. The disjunct-recovery algorithm will be successful when for all $i \in [N] \setminus S$,
$$
\sum_{j \in S} \Big(w - \frac{\dist(\bfc_i,\bfc_j)}{2}\Big) <w.
$$
Hence the condition of the lemma is sufficient for success.
\end{IEEEproof}

Now we are ready to prove Thm.~\ref{thm:indep}.
\begin{IEEEproof}[Proof of Thm.~\ref{thm:indep}] 
First of all, by union bound, 
\begin{align*}
\Pr\Big(&\exists i \in [N] : \sum_{j =1, j \ne i}^N X_j\Big(w - \frac{\dist(\bfc_i,\bfc_j)}2\Big) \ge  w\Big)\\
& \le \sum_i \Pr\Big( \sum_{j =1, j \ne i}^N X_j\Big(w - \frac{\dist(\bfc_i,\bfc_j)}2\Big) \ge w\Big). 
\end{align*}
For a fixed $i$, we would want to upper bound the probability  above in the right hand side under the summation.
Assume, $w - \frac{\dist(\bfc_i,\bfc_j)}2= a_j$. 
Notice, $ a_j X_j - \avg (a_jX_j) \le a_j(1-t/N) \le (1-t/N) (w -d/2)$ and $\sum_{j\ne i} \avg(a_jX_j - a_j t/N)^2 = \frac{t}{N}\Big(1-\frac{t}{N}\Big)\sum_{j\ne i} a_j^2.$

We have,
\begin{align*}
\Pr&\Big( \sum_{j =1, j \ne i}^N X_j \Big(w - \frac{\dist(\bfc_i,\bfc_j)}2\Big) \ge w\Big)\\
 &= \Pr\Big( \sum_{j =1, j \ne i}^N (X_j-t/N)a_j \ge w - \frac{t}{N}\sum_{j\ne i} a_j\Big)\\
&\stackrel{(a)}{\le} \Pr\Big( \sum_{j =1}^N (X_j-t/N)a_j \ge w - \frac{t}{N}\sum_{j=1}^N a_j\Big),
\end{align*}
where (a) is true as the event within the probability in second line implies the event in the third line.

Now, we can use the  classical Bernstein concentration inequality (see the version we use in  \cite[Thm.~2.7]{mcdiarmid1998concentration}), to have,
\begin{align*}
 - &\ln \Pr\Big( \sum_{j =1, j \ne i}^N X_j \Big(w - \frac{\dist(\bfc_i,\bfc_j)}2\Big) \ge w\Big)\\
& \ge  \frac{\Big(w - \frac{t}{N}\sum_{j} a_j\Big)^2}{2\Big(1-\frac{t}{N}\Big)\Big(\frac{t}{N}\sum_{j} a_j^2+\frac{1}{3}\Big(w-\frac{d}{2}\Big)\Big(w - \frac{t}{N}\sum_{j} a_j\Big) \Big)}\\
& \ge \frac{\Big(w - \frac{t}{N}\sum_{j} a_j\Big)^2}{2\Big(\frac{t}{N}\sum_{j} a_j^2+\frac{1}{3}\Big(w-\frac{d}{2}\Big)\Big(w - \frac{t}{N}\sum_{j} a_j\Big) \Big)} \\
&\ge \frac{\Big(w - \frac{t}{N}\sum_{j} a_j\Big)^2}{2\Big(\frac{t}{N}\sum_{j} a_j \Big(w-\frac{d}{2}\Big)+\frac{1}{3}\Big(w-\frac{d}{2}\Big)\Big(w - \frac{t}{N}\sum_{j} a_j\Big) \Big)} \\
&\ge  \frac{3\Big(w - \frac{t}{N}\sum_{j} a_j\Big)^2}{2\Big(w-\frac{d}{2}\Big)\Big(\frac{2t}{N}\sum_{j} a_j +w  \Big)}\\
&\stackrel{(b)}{\ge} \frac{3\Big(w - t(w-D/2)\Big)^2}{2\Big(w-\frac{d}{2}\Big)\Big(2t(w-D/2) +w  \Big)},
\end{align*}
 and (b) follows because the exponent above is an increasing function of $\sum_j a_j$ and
 $\frac{1}{N} \sum_j a_j  = w -\frac1{2N} \sum_j \dist(\bfc_i,\bfc_j) \le w - \frac{D}{2}$. 
Now using union bound, we deduce that the test matrix will successfully identify the defective elements exactly with probability $1-\epsilon$ if
\begin{align*}
\frac{3\Big(w - t(w-D/2)\Big)^2}{2\Big(w-\frac{d}{2}\Big)\Big(2t(w-D/2) +w  \Big)} \ge \ln \frac{N}{\epsilon},
\end{align*}
which proves the theorem.
\end{IEEEproof}

Similar result can be obtained for Model 2. However, because of the dependence among the random choice of defectives
we need to use  concentration inequalities for sampling without replacements.

\section{Model 2: $(t,\epsilon)$-disjunct matrices from constant-weight codes}\label{sec:proof1}

Our main result of this section is the following.
\begin{theorem}[Model 2]\label{thm:main1}
Suppose, we have a constant-weight binary code $\cC$ of size $N$, minimum distance $d$ and average distance $D$ such that every codeword has length $M$ and weight $w$. The test matrix obtained from the
code is  $(t,\epsilon)$-disjunct for the largest $t$ such that,
\begin{align*}
d \ge D - \frac{3\Big(w - t(w-D/2)\Big)^2}{ \ln \frac{N}{\epsilon}\Big(2t(w-D/2) +w  \Big)},
\end{align*}
holds. 
\end{theorem}

One can compare the results of Prop.~\ref{prop:disj} and Theorem~\ref{thm:main1}
to see the improvement achieved as we relax the definition of disjunct matrices. Indeed, 
Theorem~\ref{thm:main1} implies, $$
t  \le \frac{w -  \sqrt{\frac13(D-d)\ln \frac{N}{\epsilon}(2t(w-D/2)+w)}}{w-D/2},
$$
as opposed to $t \le \frac{w-1}{w-d/2}$ from Prop.~\ref{prop:disj}.
This will lead to the final improvement on the parameters of Porat-Rothschild
construction \cite{porat2008explicit}, as we will see in Section~\ref{sec:construction}.

\subsection{Proof of Theorem \ref{thm:main1}}
This section is dedicated to the proof of  Theorem \ref{thm:main1}.
Suppose, we have a constant-weight binary code $\cC$ of size $N$ and  minimum distance $d$ such 
that every codeword has  length $M$ and weight $w$.  Let the average distance of the code be $D.$ 
Note that this code is fixed:
we will prove the almost-disjuctness  property of this code.

Let us now choose $t$ codewords randomly and uniformly from all possible $\binom{N}{t}$ choices.
Let the randomly chosen codewords  be $\{\bfc_1,\bfc_2,\dots, \bfc_{t}\}.$
In what follows, we adapt the proof of Prop.~\ref{prop:disj} in a probabilistic setting.

Assume we call the random set of defectives as $S$.  For $l \in [N] \setminus S$,
define the
random variables
$
Z^{l} =  \mathop{\sum_{j = 1}^{t}} \Big(w - \frac{\dist(\bfc_l, \bfc_j)}2\Big).
$
Clearly, $Z^l$ is the maximum possible size of the portion of the support of $\bfc_l$ that
is common to  at least one of $\bfc_j, j =1,\dots,t.$ Note that the size of support of $\bfc_l$
is $w$.
Hence, as we have seen in the proof of Prop.~\ref{prop:disj}, if $Z^l$ is
less than $w$ for all $l$'s that are not part of the defective set,  then the disjunct
decoding algorithm will be successful.
 Therefore, we aim to find the probability
$\Pr(\exists l \in [N]\setminus S : Z^l \ge w)$  and show it to be bounded above
by $\epsilon$ under the condition of the theorem.

As the variable $Z^l$s are identically distributed, 
using union bound,   
$$
\Pr(\exists l \in [N]\setminus S : Z^l \ge w) \le (N-t) \Pr(Z^l \ge w),
$$
where  $l$ can now assumed to be uniformly distributed in $[N]\setminus S$.
In the following, we will find an upper bound on $ \Pr(Z^l \ge w).$

In \cite{mazumdar2012almost}, an upper bound on $ \Pr(Z^l \ge w)$ was found by Azuma's inequality. It turns out that by using a trick 
from Hoeffding \cite{hoeffding1963probability}, and using the Bernstein inequality we can achieve a tighter bound. First note that, 

$$
Z^l =  \sum_{j = 1}^{t} \Big(w - \frac{\dist(\bfc_l, \bfc_j)}2\Big),
$$
where, $\bfc_1,\dots,\bfc_t, \bfc_l$ are randomly and uniformly chosen $(t+1)$ codewords
from all possible $\binom{N}{t+1}$ choices.

Given, $\bfc_l$, the other codewords are randomly sampled from the code without replacement. It follows from \cite[Theorem 4]{hoeffding1963probability}, for any real number s that,
$$
\avg \Big(e^{sZ^l} \mid \bfc_l\Big) \le \avg \Big(e^{s\sum_{j = 1}^{t} \Big(w - \frac{\dist(\bfc_l, \bfx_j)}2\Big)} \mid \bfc_l\Big),
$$
where $\bfx_1,\dots,\bfx_t$ are codewords randomly and uniformly sampled from the code with replacement.
Therefore, 
$$
\avg \Big(e^{sZ^l}  \Big) \le \avg \Big(e^{s\sum_{j = 1}^{t} \Big(w - \frac{\dist(\bfc_l, \bfx_j)}2\Big)}  \Big),
$$
where $\bfc_l$ is a randomly and uniformly chosen codeword and $\bfx_1,\dots,\bfx_t$ are codewords randomly and uniformly sampled from the code $\cC\setminus\{\bfc_l\}$ 
with replacement.

Therefore, for any $s>0$, using Markov inequality,
$$
\Pr\Big(Z^l \ge w\Big) \le \avg e^{-sw}\Big(e^{s\sum_{j = 1}^{t}Y_j}  \Big),
$$
where, $Y_i \equiv w - \frac{\dist(\bfc_l, \bfx_i)}2, i=1, \dots t,$ are independent random variables with,
$$
\avg Y_i \le w - \frac{D}{2} .
$$
and $$ \avg Y_i^2 \le \Big( w - \frac{D}{2}\Big)  \Big( w - \frac{d}{2}\Big) ,$$
since $Y_i$ is a nonnegative random variable.
Now, since $Y_i$s are all independent, we can use  \cite[Thm.~2.7]{mcdiarmid1998concentration} (or its method of proof) again, to upper bound
large deviation for the sum $Z^l$. Indeed, we must have,
\begin{align*}
\Pr\Big(Z^l \ge w\Big)
& \le \exp\Big(-  \frac{(w -\sum_{i=1}^t \avg Y_i)^2}{\cA}\Big),
\end{align*}
where,
\begin{align*}
\cA &= 2\sum_{i=1}^t (\avg Y_i^2- (\avg Y_i)^2) \\
& \quad + \frac{2}{3}(w -\sum_{i=1}^t \avg Y_i) (w - \frac{d}2 - w + \frac{D}{2})\\
&\le 2t ( w - \frac{D}{2})  ( w - \frac{d}{2})- 2t (w-\frac{D}{2})^2 \\
& \quad + \frac{1}{3}(w -t (w-\frac{D}{2})) (D -d) \\
&= t ( w - \frac{D}{2})  ( D-d) + \frac{1}{3}(w -t (w-\frac{D}{2})) (D -d).
\end{align*}
Hence, we have,
\begin{align*}
\Pr\Big(Z^l \ge w\Big)
\le \exp\Big(-  \frac{3(w -t (w-\frac{D}{2}))^2}{(2t ( w - \frac{D}{2})   + w ) (D -d)} \Big).
\end{align*}

Now using union bound, we deduce that the test matrix will successfully identify the defective elements exactly with probability $1-\epsilon$ if
\begin{align*}
\frac{3\Big(w - t(w-D/2)\Big)^2}{\Big(D-d\Big)\Big(2t(w-D/2) +w  \Big)} \ge \ln \frac{N}{\epsilon},
\end{align*}
which proves the theorem.

\remove{

Define,
$$
Z_i =\avg\Big( \sum_{j = 1}^{t} \Big(w - \frac{\dist(\bfc_l, \bfc_j)}2\Big) \mid  {\dist(\bfc_l, \bfc_k)}, k=1,2,3,\dots, i  \Big).
$$
Clearly,
$
Z_0 = \avg \Big( \sum_{j = 1}^{t} \Big(w - \frac{\dist(\bfc_l, \bfc_j)}2\Big) \Big),
$
and
$
Z_{t} =  \sum_{j = 1}^{t} \Big(w - \frac{\dist(\bfc_l, \bfc_j)}2\Big) = Z^l.
$

Now,
\begin{align*}
Z_0 &= \avg \Big( \sum_{j = 1}^{t} \Big(w - \frac{\dist(\bfc_l, \bfc_j)}2\Big) \Big)
= tw - \frac12 \avg \sum_{j = 1}^{t}\dist(\bfc_l, \bfc_j),
\end{align*}
where the expectation is over the randomly and uniformly chosen $(t+1)$ codewords
from all possible $\binom{N}{t+1}$ choices. Note,
\begin{align*}
&\avg \sum_{j = 1}^{t}\dist(\bfc_l, \bfc_j) =  \sum_{i_1<i_2<\dots<i_{t+1}}\frac1{\binom N{t+1}}\sum_{j=2}^{t+1} \dist(\bfc_{i_1}, \bfc_{i_j})\\
& =\frac1{(t+1)!\binom N{t+1}}\mathop{\sum_{i_l\ne i_m}}_{1\le l \ne m\le t+1} \sum_{j=1}^{t}
\dist(\bfc_{i_1}, \bfc_{i_j})=\frac1{N(N-1)}\sum_{j=2}^{t+1}\sum_{i_1=1}^N\sum_{i_j\ne i_1}
 \dist(\bfc_{i_1}, \bfc_{i_j})\\
&=\sum_{j=2}^{t+1}\avg \dist(\bfc_{i_1}, \bfc_{i_j}) \ge tD,
\end{align*}
where the  expectation on the last line is over a uniformly chosen pair of distinct random codewords
of $\cC.$ Hence,
$$
Z_0  \le t(w-D/2).
$$

We start with the lemma below.
\begin{lemma}\label{lem:marti}
The sequence of random variables $Z_i, i =0, 1, \dots , t,$ forms a martingale.
\end{lemma}
The statement is true by construction. For completeness we present a  proof that is deferred to Appendix \ref{app:one}.
Once we have proved that the sequence is a martingale, we show that it is a bounded-difference
martingale.
\begin{lemma}\label{lem:bounded}
For any $i = 1,\dots,t$,
$$
|Z_i -Z_{i-1}| \le (w-d/2) \Big(  1 +  \frac{t-i}{N-i}\Big).
$$
\end{lemma}
The proof is deferred to Appendix \ref{app:two}.

Now using Azuma's inequality for martingale with bounded difference \cite{mcdiarmid1989method},
we have,
$$
\Pr\Big(Z_{t} -Z_0 \ge  \nu) \le \exp\Big(-\frac{\nu^2}{2(w-d/2)^2\sum_{i=1}^{t}c_i^2} \Big),
$$
where, $c_i =   1 +  \frac{t-i}{N-i}$.
This implies,
$$
\Pr\Big(Z_{t} \ge \nu + t(w-D/2)\Big) \le \exp\Big(-\frac{\nu^2}{2(w-d/2)^2\sum_{i=1}^{t}c_i^2} \Big).
$$
Setting, $\nu = w -  t(w-D/2)$, we have,
\begin{align*}
\Pr\Big(Z^l \ge w\Big)
&\le \exp\Big(-\frac{(w -  t(w-D/2))^2}{2(w-d/2)^2\sum_{i=1}^{t}c_i^2} \Big).
\end{align*}
Now,
\begin{align*}
\sum_{i=1}^{t}c_i^2 &\le t\Big(1+\frac{t-1}{N-1}\Big)^2.
\end{align*}
Hence,
\begin{align*}
\Pr(\exists i \in [N]\setminus S : Z^i \ge w) &\le (N-t)\exp\Big(-\frac{(w -  t(w-D/2))^2}{2t(w-d/2)^2\Big(1+  \frac{t-1}{N-1}\Big)^2} \Big)< \epsilon,
\end{align*}
when,
$$
d/2 \ge w - \frac{w-t(w-D/2)}{\alpha\sqrt{t\ln\frac{N-t}{\epsilon}}},
$$
and $\alpha$ is a constant greater than $\sqrt{2}\Big(1+  \frac{t-1}{N-1}\Big).$

}

\subsection{Higher order statistics of distance distribution}
We get slightly tighter bounds in both the Theorems \ref{thm:indep} and \ref{thm:main1}, if higher order than only the average distance 
of the codes have been considered. 
Indeed in both of the main theorems 
 we have  used the
inequality,
$$
 \frac{1}{|\cC|^2} \sum_{\bfy, \bfx \in \cC} (w - \frac{\dist(\bfx,\bfy)}{2})^2 \le (w- \frac{d}{2})  \frac{1}{|\cC|^2} \sum_{\bfy, \bfx \in \cC} (w - \frac{\dist(\bfx,\bfy)}{2}),
$$
since $w -\frac{\dist(\bfx,\bfy)}{2}$ is always nonnegative. However both of the theorems could be rephrased
in terms of the second-moment of the distance distribution. For example, Theorem  \ref{thm:main1} can be restated with slightly stronger result.
\begin{theorem}\label{thm:d2}
Suppose, we have a constant-weight $(M, N, d, w)$ binary code $\cC$ with  average distance $D$ and the second-moment of the
distance distribution $D_2$. The test matrix obtained from the
code is  $(t,\epsilon)$-disjunct for the largest $t$ such that,
\begin{align}\label{eq:d2}
d \ge D + \frac{3t(D_2- D^2)}{2\Big(w - t(w-D/2)\Big)}- \frac{3\Big(w - t(w-D/2)\Big)}{ \ln \frac{N}{\epsilon}},
\end{align}
holds. 
\end{theorem}
We omit the proof of this theorem as it is exactly same as the proof of Theorem \ref{thm:main1}.

However, it turns out (in the next section) that our results, that rely only on the average distance, are sufficient to give
near-optimal performance in group testing schemes in terms of the number of tests. In particular, use of \eqref{eq:d2}
in conjunction with the construction of constant-weight codes below, instead of  Theorem \ref{thm:main1}, leads to improvement only on the constant terms.

\section{Construction}\label{sec:construction}
\subsection{Discussions}
As we have seen in Section~\ref{sec:disjunct}, constant-weight codes can be used to
produce disjunct matrices. Kautz and Singleton \cite{kautz1964nonrandom} gives a construction
of constant-weight codes that results in good disjunct matrices.
In their construction, they start with a Reed-Solomon (RS) code, a
$q$-ary error-correcting code  of length
$q-1.$ For a detailed discussion of RS codes we refer the reader
to the standard textbooks of coding theory \cite{MS1977,roth2006introduction}.
Next they replace the $q$-ary symbols in the codewords by unit weight
binary vectors of length $q$. The mapping from $q$-ary symbols to length-$q$
unit weight binary vectors is bijective: i.e., it is $0\to 100\dots0; 1\to 010\dots0;\dots;
q-1\to 0\dots01.$ We refer to this mapping as $\phi.$ As a result, one obtains a set of  binary vectors
of length $q(q-1)$ and constant-weight $q.$ The size of the resulting binary code
is same as the size of the RS code, and the distance of the binary code
is twice that of the distance of the RS code.

For a $q$-ary RS code of size $N$ and length $q-1$, the
minimum distance is $q-1-\log_q{N}+1 = q-\log_q{N}.$ Hence,
the Kautz-Singleton construction is a constant-weight code with length $M=q(q-1)$, weight $w=q-1$,
size $N$ and distance $2(q-\log_q{N})$. Therefore,
from Prop.~\ref{prop:disj},
we have a $t$-disjunct matrix with,
\begin{align*}
t =& \frac{q-1-1}{q-1-q+\log_q{N}} = \frac{q-2}{\log_q{N}-1} \\
& \qquad \approx \frac{q\log q}{\log N} \approx \frac{\sqrt{M}\log M}{2\log N}.
\end{align*}
On the other hand, note that, the average distance of the RS code is
$(q-1)(1-1/q).$ Hence the average distance of the resulting
constant-weight code from Kautz-Singleton construction will be
$$
D= \frac{2(q-1)^2}{q}.
$$
Now, substituting these values in Theorem~\ref{thm:main1}, we have a 
$(t,\epsilon)$ disjunct matrix, where,
\begin{align*}
2(q &- \log_q N) \\
& \ge  2\frac{(q-1)^2}{q} 
- \frac{3(q-1 - t(q-1- \frac{(q-1)^2}{q} ))^2}{(2t(q-1 - \frac{(q-1)^2}{q})+q-1)\ln \frac{N}{\epsilon}}\\
& = \frac{2(q-1)^2}{q} - \frac{3(q-1)(1-t/q)^2}{(1+2t/q)\ln \frac{N}{\epsilon}}.
\end{align*}

This basically restricts $t$ to be about $O(\sqrt{M})$ (since, $ 1 - t/q$ must be nonnegative). Hence, Theorem~\ref{thm:main1} does not
obtain any meaningful improvement from the Kautz-Singleton construction in the asymptotics except in special cases.

There are two places where the Kautz-Singleton construction can be modified:
1) instead of Reed-Solomon code one can use any other
$q$-ary code of different length, and 2) instead of the mapping $\phi$
any binary constant-weight code  of size $q$ might have been used.
For a general discussion we refer the reader to \cite[\S7.4]{du1993combinatorial}.
In the recent work \cite{porat2008explicit}, the mapping $\phi$ is kept the same, while
the RS code has been changed to a $q$-ary code that achieve the Gilbert-Varshamov
bound \cite{MS1977, roth2006introduction}.

In our construction of disjunct matrices we use the Kautz-Singleton construction and instead of Reed-Solomon code either
1) follow the footsteps of \cite{porat2008explicit} to use a Gilbert-Varshamov code or 2) use Algebraic-Geometric codes.  We exploit some  property
of the resulting scheme (namely, the average distance) and do a
finer analysis that was absent from the previous works such as \cite{porat2008explicit}.

\subsection{$q$-ary Gilbert-Varshamov  construction}
Next, we construct a {\em linear} $q$-ary code
of size $N$, length $M_q$ and minimum distance $d_q$ that
achieves the Gilbert-Varshamov (GV) bound \cite{MS1977,roth2006introduction}. We describe the bound in Appendix \ref{sec:explicit}.

Porat and Rothschild \cite{porat2008explicit} show that it is
possible to construct  in time $O(M_q N)$ a $q$-ary code that achieves the GV
bound. To have such construction, they exploit the
following  well-known fact: a $q$-ary linear code with random generator matrix
achieves the  GV bound with high probability \cite{roth2006introduction}. To have an explicit construction
of such codes, a derandomization method known as the method of conditional
expectation \cite{AS2000} is used. In this method, the entries of the generator
matrix of the code are chosen one-by-one so that the minimum distance of the resulting code
does not go below the value prescribed by Eq.~(\ref{eq:GV}). For a detail description of
the procedure, see \cite{porat2008explicit}.

Using the GV code construction of Porat and Rothschild and plugging it in the Kautz-Singleton construction above, we
have the following proposition. 
\begin{proposition}\label{prop:const}
Let $s \le q$. There exists a polynomial time constructible family of $(M,N,2M/q(1-1/s),M/q)$-constant-weight binary code that satisfy,
\begin{equation}
M/q \le \frac{s\ln N}{\ln (q/s) -1}.
\end{equation}
\end{proposition}
Although the proof of the above proposition is essentially in Porat and Rothschild  \cite{porat2008explicit}, we have a cleaner proof that 
we include  in Appendix  \ref{sec:explicit} for completeness. 

However, we are also concerned with the average distance of the code. Indeed, we have the following proposition.

\begin{proposition}\label{prop:avg}
The average distance of the code constructed in  Prop.~\ref{prop:const} is 
$$
D =  \frac{2M}{q}(1-1/q).
$$
\end{proposition}
\begin{IEEEproof}
For Prop.~\ref{prop:const} we have followed the Kautz-Singleton construction.
We take a linear $q$-ary code $\cC'$ of length $M_q \triangleq \frac{M}{q}$, size $N$ and
minimum distance $d_q  \triangleq \frac{d}{2}.$ Each $q$-ary symbol in the codewords is then
replaced with a binary indicator vector of length $q$ (i.e., the binary vector whose all entries
are zero but one entry, which is $1$) according to the map $\phi$. As a result we have a binary code $\cC$ of length $M$ and
size $N$. The minimum distance of the code is $d$ and the codewords are of constant-weight $w = M_q = \frac{M}{q}.$
The average distance of this code is twice the average distance of the $q$-ary code. As $\cC'$ is linear
(assuming it has no all-zero coordinate), it has average distance equal to
\begin{align*}
\frac{1}{N}\sum_{j =1}^{M_q} j A_j &= \frac{N}{N}\sum_{j=0}^{M_q}j\binom{M_q}{j}(1-1/q)^j(1/q)^{M_q-j}\\
& =  M_q(1-1/q),
\end{align*}
where $A_j$ is the number of codewords of weight $j$ in $\cC'.$
Here we use the fact that the average of the  distance between any two randomly chosen codewords of
a nontrivial linear code is equal to that of a binomial random variable  \cite{MS1977}.
Hence the constant-weight code $\cC$ has average distance
$D =  2M_q(1-1/q).$
\end{IEEEproof}
\subsection{Constructions for Model 1}
We follow the Kautz-Singleton  code construction. Suppose, we have a $(M,N, d, M/q)$- constant-weight code 
that satisfies Prop.~\ref{prop:const} and \ref{prop:avg}.
Hence,  average distance $D =  \frac{2M}{q}(1-1/q)$.
The resulting test matrix will satisfy the condition of Thm.~\ref{thm:indep} when,

\begin{equation}
d/2 \ge  M/q  -  \frac{3\Big(M/q - t(M/q-M/q(1-1/q))\Big)^2}{2 \Big(2t(M/q-M/q(1-1/q)) +M/q  \Big)\ln \frac{N}{\epsilon}}
\end{equation}
or when,
\begin{equation}\label{eq:inter}
d/2 \ge  M/q  -  \frac{3M/q\Big(1 - t/q\Big)^2}{2 \Big(2t/q +1  \Big)\ln \frac{N}{\epsilon}}.
\end{equation}
Hence a sufficient condition is to chose the constant-weight  code such that,
$$
d \ge \frac{2M}{q}\Big(1 - \frac{3\Big(1 - t/q\Big)^2}{2 \Big(2t/q +1  \Big)\ln \frac{N}{\epsilon}}\Big).
$$
We can take $q$ to be the smallest power of prime that is greater than $2t$. Which will make the sufficient condition look like,
$$
d\ge  \frac{2M}{q}\Big(1 - \frac{3}{16\ln \frac{N}{\epsilon}}\Big).
$$
However, according to Prop.~\ref{prop:const}, such code can be explicitly constructed with,
\begin{equation}
M/q \le \frac{16/3 \ln \frac{N}{\epsilon} \ln N}{\ln (3t/(16 \ln \frac{N}{\epsilon})) -1}.
\end{equation} 
Hence, the sufficient number of tests is $M  = \frac{6t}{\ln t} \ln N \ln \frac{N}{\epsilon}.$



\subsection{Construction of almost disjunct matrix: Model 2}
We again follow the above code construction and
 choose $q$ to be a power of a prime number. 
With proper parameters we can have a disjunct matrix with
the following property.
\begin{theorem}\label{thm:main2}
 It is possible to explicitly construct a $(t,\epsilon)$-disjunct matrix of size $M\times N$ where
$$
M  = O\Big(\frac{t}{\log t} \log N \log \frac{N}{\epsilon}\Big).
$$
\end{theorem}

\begin{IEEEproof} 
We follow the Kautz-Singleton  code construction as earlier. That is we have a $(M,N, d, M/q)$- constant-weight code 
that satisfies Prop.~\ref{prop:const} and \ref{prop:avg}.
Hence,  average distance $D =  \frac{2M}{q}(1-1/q)$.
The resulting matrix will be $(t,\epsilon)$-disjunct if the condition of Theorem \ref{thm:main1} is satisfied,
i.e.,

\begin{align*}
d & \ge  \frac{2M}{q}(1-1/q)\\
&\qquad - \frac{3\Big(M/q - t(M/q-M/q(1-1/q))\Big)^2}{ \Big(2t(M/q-M/q(1-1/q)) +M/q  \Big)\ln \frac{N}{\epsilon}},
\end{align*}

or when,
\begin{equation}
d \ge  \frac{2M}{q}(1-1/q)  -  \frac{3M/q\Big(1 - t/q\Big)^2}{ \Big(2t/q +1  \Big)\ln \frac{N}{\epsilon}}.
\end{equation}
Hence a sufficient condition is to choose the constant-weight  code such that,
$$
d \ge \frac{2M}{q}\Big(1 - \frac1q - \frac{3\Big(1 - t/q\Big)^2}{2 \Big(2t/q +1  \Big)\ln \frac{N}{\epsilon}}\Big).
$$
Since the requirement of above sufficient condition is slightly weaker than  that of \eqref{eq:inter}, we can 
still choose $q$ to be a smallest power of prime that is greater than $2t$, and follow the calculations for Model 1,  to
\remove{
We can chose $q$ to be a smallest power of prime that is greater than $2t$. Which will make the sufficient condition look like,
$$
d\ge  \frac{2M}{q}\Big(1 - \frac{3}{16\ln \frac{N}{\epsilon}}\Big).
$$
However, according to Prop.~\ref{prop:const}, such code can be explicitly constructed with,
\begin{equation}
M/q \le \frac{16/3 \ln \frac{N}{\epsilon} \ln N}{\ln (3t/(16 \ln \frac{N}{\epsilon})) -1}.
\end{equation} 
Hence,}
obtain 
total number of tests  $M  = O\Big(\frac{t}{\ln t} \ln N \ln \frac{N}{\epsilon}\Big).$

\end{IEEEproof}
\vspace{0.1in}

It is clear from Prop.~\ref{prop:scheme} that a $(t,\epsilon)$ disjunct
matrix is equivalent to a group testing scheme. Hence, as
 a consequence of Theorem~\ref{thm:main2},  
we will be able to construct a testing scheme with $O\Big(\frac{t}{\log t} \ln N \ln \frac{N}{\epsilon}\Big)$
tests. Whenever the defect-model is such that all the possible defective sets of
size $t$ are equally likely and there are no more than $t$ defective elements,
the above group testing scheme will be successful with probability at lease $1-\epsilon.$

Note that, if $t$ is proportional to any positive power of $N$, then $\log N$ and
$\log t$ are of same order. Hence it will be possible to have the above testing
scheme with $O(t\log \frac{N}{\epsilon})$ tests, for any  $\epsilon> 0$.

\subsection{Constructions based on Algebraic-Geometric codes}
Now, instead of using the Porat-Rothschild construction of GV codes, we can use the
 Algebraic-Geometric (AG) code construction of  Tsfasman, Vl\u{a}du\c{t} and Zink \cite{tsfasman1982modular}.
 In particular, we can base our construction on the  Garc\'{i}a-Stichtenoth Tower of function field over $\ff_q$ \cite[Sec.~3.4.3]{tsfasman2007algebraic}.

Assume, $q = r^2$, where $r$ is any integer. For any even number $n$, there is a family of modular curves with genus $g_n = (r^{n/2}-1)^2$ with number of
points given by $M_q \ge r^{n+1} - r^n +1$ (see, \cite[Theorem 3.4.44]{tsfasman2007algebraic}). Now, using Corollary 4.1.14 of \cite{{tsfasman2007algebraic}}, we conclude
that it is possible to construct families of linear code of length $M_q$, size $N$ and minimum distance $d_q$, where,
$$
M_q \ge r^{n+1} - r^n +1,
$$
and
$$
\log_q N  = M_q - d_q -g_n +1.
$$
Hence, we obtain families of linear code such that,
\begin{align}
\frac{\log_q N}{M_q} \ge 1 - \frac{d_q}{M_q} - \frac{1}{\sqrt{q}-1}.
\end{align}
Now, using the Kautz-Singleton mechanism of converting this to a binary code, we obtain an $(M,N,d,w)$ constant weight code, where
$$
M =q M_q; d = 2d_q; w = M_q = M/q,
$$   
and,
\begin{align}\label{eq:9}
d \ge \frac{2M}{q}\Big(1 - \frac{q\log_q N}{M}  - \frac{1}{\sqrt{q}-1}\Big).
\end{align}
 Since, the  AG code is a linear code, we can calculate the 
the average distance of the above constant-weight code as in Proposition~\ref{prop:avg}. Indeed,
the average distance $D = \frac{2M}{q}(1-\frac{1}{q})$.

For this family of codes, we can also calculate the second-moment of the distance distribution\footnote{It turns out that $D_2 = D^2+ \frac{4M}{q^2}(1-\frac{1}{q})$.}, that allows us to use Theorem~\ref{thm:d2}.
To be consistent of the rest of the paper, we rely on only the average distance, and use Theorem~\ref{thm:main1} instead. Substituting the values of $D, w$ in Theorem~\ref{thm:main1}, we obtain the following. If 
\begin{align}\label{eq:10}
d \ge \frac{2M}{q}\Big(1 - \frac1q - \frac{3\Big(1 - t/q\Big)^2}{2 \Big(2t/q +1  \Big)\ln \frac{N}{\epsilon}}\Big),
\end{align}
then the construction is $(t,\epsilon)$-almost disjunct. Comparing \eqref{eq:9} and \eqref{eq:10}, we claim that, our construction is 
 $(t,\epsilon)$-almost disjunct as long as,
 \begin{align}
 \frac{q\log_q N}{M}  + \frac{1}{\sqrt{q}-1} \le  \frac1q + \frac{3\Big(1 - t/q\Big)^2}{2 \Big(2t/q +1  \Big)\ln \frac{N}{\epsilon}}.
\end{align}
Now assuming $q$ to be the smallest power of $2$ greater than $2t$, we see that the above condition is satisfied when,
$$
M \ge \frac{16t\ln N}{\ln 2t}\ln \frac{N}{\epsilon}.
$$
We should note that construction for Model 1 can be done in the exact same way to obtain the same parameters. 

\vspace{0.2in}

\begin{remark} {\em (The traditional argument (Prop.~\ref{prop:disj}) with Algebraic-Geometric Codes)}
Note that, one could use AG codes in conjunction with Prop.~\ref{prop:disj} to obtain disjunct matrices. However
such a construction results in highly suboptimal number of rows (tests). Indeed, substituting Eq.~\eqref{eq:9} in Prop.~\ref{prop:disj},
we have a $t$-disjunct matrix with,
$$
t = \frac{1}{\frac{q\log_q N}{M}  + \frac{1}{\sqrt{q}-1}} \Rightarrow \frac{q\log_q N}{M} = \frac1t - \frac{1}{\sqrt{q}-1}.
$$
Hence, to get anything nontrivial we must have $q \ge t^2$, which results in $M = \Omega(t^3 \log N/\log t)$. This is quite bad compared to 
the optimal constructions that give disjunct matrices with $O(t^2 \log N)$ rows. It is interesting that by using our average distance based 
arguments we are able to get rid of such suboptimality with AG codes. Intuitively, while the range of minimum distance of the constant-weight codes
obtained from the AG codes is not sufficient for optimal results, the combination of average distance and minimum distance for these codes indeed 
belongs to the best possible region.
\end{remark}

\section{Conclusion}
In this work we show that it is possible to construct non-adaptive 
group testing schemes with small number of tests that identify a
uniformly chosen random defective configuration with high probability.
To construct a $t$-disjunct matrix 
one starts with the simple relation between the minimum distance $d$ of a
constant $w$-weight code and $t.$ This is an example of a scenario where
a pairwise property (i.e., distance) of the elements of a set is
translated into a property of $t$-tuples.

Our method of analysis provides a general way to prove that a property
holds for almost all $t$-tuples of elements from a set based on the mean pairwise statistics
of the set. Our method might be useful in many areas of applied combinatorics,
such as digital fingerprinting or design of
key-distribution schemes, where such a translation is evident. With this method
potential new results may be obtained for
 the cases of cover-free codes \cite{d2002families,stinson2000some,kautz1964nonrandom}, traceability and frameproof
codes \cite{chor1994tracing,staddon2001combinatorial}.

\appendix

\subsection{Gilbert-Varshamov bound and proof of Prop.~\ref{prop:const}}\label{sec:explicit}
\begin{lemma}[Gilbert-Varshamov Bound]
There exists an $(m, N, d)_q$-code such that,
\begin{equation}\label{eq:GV}
N \ge \frac{q^m}{\sum_{i=0}^{d-1} \binom{m}{i}(q-1)^i}.
\end{equation}
\end{lemma}
\begin{corollary}\label{cor:gv}
Suppose $X$ is a Binomial($m, 1-\frac{1}{q}$) random variable. There exists an $(m, N, d)_q$-code such that,
$$
N \ge \frac{1}{\Pr(X\le d)}.
$$
\end{corollary}
\begin{lemma}\label{lem:ld}
Suppose $X$ is a Binomial($m, 1-\frac{1}{q}$) random variable. Then, for all $s<q$,
\begin{equation}
\Pr\Big(X \le m\Big(1-\frac{1}{s}\Big)\Big) \le e^{-mD(1/s || 1/q)},
\end{equation}
where $D(p||p') = p \ln(p/p') +(1-p)\ln((1-p)/(1-p')).$ 
\end{lemma}
\begin{theorem}\label{thm:gv_exist}
Let $s <q$. For the $(m, N, m(1-1/s))_q$-code that achieves the Gilbert-Varshamov bound, we have
\begin{equation}
m \le \frac{s\ln N}{\ln (q/s) -1}.
\end{equation}
\end{theorem}
\begin{IEEEproof}
This theorem follows from corollary \ref{cor:gv} and lemma \ref{lem:ld}. Note that,
\begin{align*}
D(1/s || 1/q) & = \frac1s \ln \frac{q}{s} + \Big(1-\frac1s\Big) \ln \Big(1-\frac1s\Big) \\
& \qquad- \Big(1-\frac1s\Big)\ln\Big(1-\frac1q\Big)\\
& \ge  \frac1s \ln \frac{q}{s} + \Big(1-\frac1s\Big) \ln \Big(1-\frac1s\Big)\\
&\ge   \frac1s \ln \frac{q}{s} -\frac1s,
\end{align*}
where in the last line we have used the fact that $x\ln x \ge x-1$ for all $x >0$.
\end{IEEEproof}
Using the Kautz-Singleton construction, this implies that,
there exists a polynomial time constructible family of $(M,N,2M/q(1-1/s),M/q)$-constant-weight binary code with,
$$
M/q \le \frac{s\ln N}{\ln (q/s) -1},
$$
which is Prop.~\ref{prop:const}.

\vspace{0.2in}
{\em Acknowledgements:} The author would like to thank Alexander Barg  for many discussions
related to the group testing problem.

\bibliographystyle{abbrv}
\bibliography{aryabib}
%
%

\end{document}